\documentclass[a4paper]{jpconf}
\usepackage{graphicx}
\def\degC{\kern-.2em\r{}\kern-.3em C }

\begin{document}
\title{Spontaneous magnetization and magnetic susceptibility of LaVO$_{3}$
}

\author{Hiroya Sakurai}

\address{National Institute for Materials Science, 1-1 Namiki, Tsukuba 305-0044, Japan}

\ead{sakurai.hiroya@nims.go.jp}

\begin{abstract}
Five characteristic temperatures of $T_{\mathrm{M}} = 148$ K, $T_{\mathrm{N}} = 142$ K, $T_{\mathrm{t}} = 138$ K, $T_{\mathrm{f}} \sim 125$ K and $T_{\mathrm{g}} \sim 50$ K were found by the measurements of the magnetization curves at various temperatures. 
The spontaneous magnetization appears below $T_{\mathrm{M}}$. It increases up to $M_{\mathrm{S}} \simeq 2 \times 10^{-4}$ $\mu_{B}$ at $T_{\mathrm{t}}$ and then decreases steeply below $T_{\mathrm{t}}$, which qualitatively agrees with the temperature dependence of magnetization obtained under field-cooling (FC) condition. 
On the other hand, the slope of the magnetization curve, namely the magnetic susceptibility, drops below $T_{\mathrm{N}}$, which coincides with the temperature dependence of magnetization obtained under zero-FC condition, although the magnetization curves were obtained under FC condition. 
The temperature dependence of the spontaneous magnetization shows a minimum at $T_{\mathrm{f}}$ and a drop at $T_{\mathrm{g}}$ although there is no anomaly in the temperature dependence of FC or ZFC magnetization.
\end{abstract}

\section{Introduction}
The perovskite oxide, LaVO$_{3}$, shows the giant diamagnetism \cite{ShirakawaJJAP, MahajanPRB}. 
The magnetization obtained under field-cooling (FC) condition becomes largely negative when the applied field is below 0.2 T or a little higher; 
the $M/H$ value at 0.02 T, for example, is about $-8 \times 10^{-3}$ emu/mol. 
The temperature dependence of magnetization under zero-FC (ZFC) condition shows the conventional behavior of an antiferromagnet with the peak at $T_{\mathrm{N}} = 142$ K \cite{GoodenoughCRAS,NguyenPRB}. 
On the other hand, FC data have an unusual enhancement around $T_{\mathrm{N}}$ and a sharp drop below $T_{\mathrm{t}} = 138$ K. 
The canted antiferromagnetic (AF) state has been confirmed below $T_{\mathrm{t}}$  \cite{MahajanPRB} and suggested between $T_{\mathrm{t}}$ and $T_{\mathrm{N}}$ \cite{NguyenPRB}. 
Obviously the enhancement is the key to understand the giant diamagnetism. 
Nevertheless, there is no report on whether the spontaneous magnetization appears between $T_{\mathrm{t}}$ and $T_{\mathrm{N}}$. 
The study on the magnetism of the compound is not sufficient although a large number of reports on it have been published.
The purpose of this paper is to show the magnetic data, which will be useful to elucidate the magnetism in future.

\section{Experiments}
The powder sample of LaVO$_{3}$ was obtained by reduction of LaVO$_{4}$ in H$_{2}$ gas at 1200\degC for overnight. 
Then, the sample was annealed in flowing Ar gas at 1200\degC for overnight. 
The oxygen content of the final product was estimated from the weight increase after firing the sample at 600\degC in air. 
The exact composition of the sample is LaVO$_{2.98\pm 0.01}$. 
LaVO$_{4}$ was prepared from the stoichiometric mixture of La$_{2}$O$_{3}$ and V$_{2}$O$_{5}$ at 600\degC for overnight and then 650\degC for overnight. 
Magnetic measurements were performed on MPMS-XL (Quantum Design). 
The magnetization curves at various temperatures were measured from 7 T to $-0.1$ T with the intervals of 0.2 T down to 1 T and of 0.05 T below 1 T. 
To obtain each magnetization curve, the sample was once heated at 160 K, and the magnetic field of 7 T was applied there.
Then the sample was cooled down under the field to a target temperature to start magnetization measurement. 
Two sets of the measurements of the magnetization curves were performed using two samples, which were taken from a single batch of the sample.
They were named Run 1 and Run 2, which agree well with each other.
In this paper, $M$, $H$, $\chi$, and $T$ represent magnetization, magnetic field, magnetic susceptibility, and temperature, respectively. 
Magnetization curve between 0.1 T and 1 T at each temperature was fit by the linear function of $M = M_{\mathrm{S}} + \chi H$, where $M_{\mathrm{S}}$ represents spontaneous magnetization.

\section{Results and Discussion}
\begin{figure}[h]
\includegraphics[width=14pc]{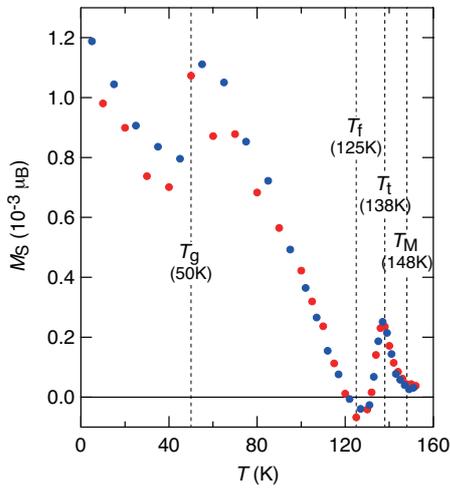}\hspace{2pc}%
\begin{minipage}[b]{14pc}\caption{\label{MsT}
Temperature dependence of spontaneous magnetization. The red and blue data were obtained by Run 1 and Run 2, respectively.
}
\end{minipage}
\end{figure}

The temperature dependence of the spontaneous magnetization is shown in Fig. \ref{MsT}. 
There are four characteristic temperatures of $T_{\mathrm{M}} = 148$ K, $T_{\mathrm{t}} = 138$ K, $T_{\mathrm{f}} \sim 125$ K, and $T_{\mathrm{g}} \sim 50$ K. 
The spontaneous magnetization appears below $T_{\mathrm{M}}$, and it develops up to $M_{\mathrm{S}} = 2.5 \times 10^{-4}$ $\mu_{B}$ at $T_{\mathrm{t}}$. 
Below $T_{\mathrm{t}}$ the spontaneous magnetization decreases steeply, but it starts to increase again at around $T_{\mathrm{f}}$. 
The small drop of the spontaneous magnetization happens at $T_{\mathrm{g}}$.
Obviously the enhancement between $T_{\mathrm{M}}$ and $T_{\mathrm{t}}$ indicates that magnetic ordering occurs at $T_{\mathrm{M}}$, not $T_{\mathrm{N}}$, because the spontaneous magnetization never appear in a paramagnetic state.
The $M_{\mathrm{S}}$ values above 148 K are constant at approximately $5 \times 10^{-5}$ $\mu_{B}$, 
which was likely caused by inevitable inaccuracy of the magnetic field.
So, the spontaneous magnetization at $T_{\mathrm{t}}$ may be estimated to be approximately $2.0 \times 10^{-4}$ $\mu_{B}$.
It should be pointed out that the temperature dependence of the spontaneous magnetization is concave or linear just below $T_{\mathrm{M}}$, which is quite unusual for conventional canted antiferromagnetic ordering.
Usually it is convex because it corresponds to the temperature dependence of the magnetization of each magnetic sublattice, which is the order parameter of the magnetic transition.

\begin{figure}[h]
\includegraphics[width=14pc]{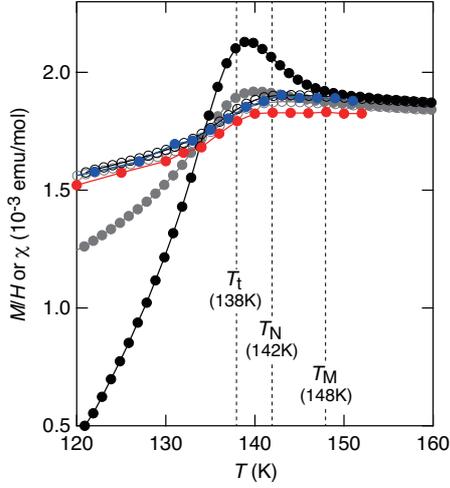}\hspace{2pc}%
\begin{minipage}[b]{14pc}\caption{\label{chiT}
Magnetization or magnetic susceptibility as a function of temperature. The open and closed markers represent ZFC and FC data, respectively. The red, blue, black and gray markers represent $\chi$ obtained by Run 1 and Run 2, and $M/H$ for 0.01 T and 1T, respectively.
}
\end{minipage}
\end{figure}

The temperature dependence of magnetic susceptibility is shown in Fig. \ref{chiT}, together with that of magnetization under FC and ZFC conditions at $H = 0.01$ T and 1 T.
It agrees well with $M/H$-$T$ curves under ZFC condition, not those under FC condition, although the magnetization curves were obtained under FC condition.
They all show the unusual peak at $T_{\mathrm{N}}$;
since the magnetic susceptibility here corresponds to the AC susceptibility at the lowest frequency, it should have the peak exactly at magnetic transition temperature \cite{SakuraiACIE}.

\begin{figure}[h]
\includegraphics[width=14pc]{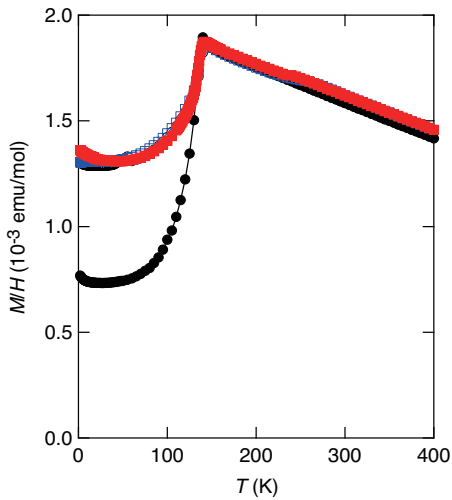}\hspace{2pc}%
\begin{minipage}[b]{14pc}\caption{\label{MT}
Temperature dependence of magnetization under 7 T (red and blue squares) and 1 T (black circles). The open and full markers represent ZFC and FC data, respectively.
}
\end{minipage}
\end{figure}

The spontaneous magnetization is close to null at around $T_{\mathrm{f}}$, which, naively thinking, implies that there are two kinds of magnetic moments canceling out each other around $T_{\mathrm{f}}$. 
Indeed, the internal field at the V site estimated by $^{57}$Fe M\"{o}ssbauer spectroscopy seems to disappear at around 115 K \cite{YoonJAP}, which is in fair agreement with $T_{\mathrm{f}}$ considering the difference between bulk magnetization and the internal field at V site, and possible sample dependence. 
Of course, it can be assumed that two kinds of magnetic moments develop independently from $T_{\mathrm{M}}$ and $T_{\mathrm{t}}$. 
In this case, the magnetic moments ordering at $T_{\mathrm{t}}$ are fixed in opposition to the magnetic field due to those ordering at $T_{\mathrm{M}}$ because the former is smaller in the magnitude just below $T_{\mathrm{t}}$ than the latter.
The former becomes larger in the magnitude below $T_{\mathrm{f}}$.
However, it is also possible to expect that two kinds of magnetic moments order simultaneously at $T_{\mathrm{M}}$;
the smaller magnetic moments may develop quickly but the larger ones slowly. 
This scenario seems to be more probable considering the unusual development of $M_{\mathrm{S}}$ just below $T_{\mathrm{M}}$.
In either case, magnetic moments can flip over at around $T_{\mathrm{f}}$ when a large magnetic field is applied, and so no significant difference was observed at $T_{\mathrm{f}}$ in the magnetization between FC and ZFC data under the field.
This can give the reason why there is no large difference between FC and ZFC magnetization under 7 T (See Fig. \ref{MT}).
The possible origins of the two kinds of magnetic moments would be the spins and the orbitals of V ions \cite{ZhouPRL}.

The anomaly at $T_{\mathrm{g}}$ corresponds to the glassy behavior of the thermal conductivity above around $T_{\mathrm{g}}$ \cite{ZhouPRL}, 
which suggests that a part of the orbital degree of freedom survives down to $T_{\mathrm{g}}$ and orders at $T_{\mathrm{g}}$ to cause the drop of the magnetization. 
The magnitude of the drop at $T_{\mathrm{g}}$ almost coincides with that at $T_{\mathrm{t}}$, supporting the similarity between the origins of the drops.
Khaliullin $et$ $al.$ suggested a partial orbital ordering and G-type spin order below the partial-orbital-ordering temperature \cite{KhaliullinPRL,KhaliullinPTP}.
Since C-type spin order has been confirmed by neutron diffraction measurement at 4.2 K \cite{ZubkovSPSS}, the type of the spin order may change at $T_{\mathrm{g}}$ although another neutron diffraction study focused on no change in the spin structure at around $T_{\mathrm{t}}$ \cite{TungPRB}. 
Because the orbital state of LaVO$_{3}$ could depend on thermal and field history of the sample, careful experiments would be needed to clarify the magnetism of the compound.

\section{Summary}
Five characteristic temperatures of $T_{\mathrm{M}}$, $T_{\mathrm{N}}$, $T_{\mathrm{t}}$, $T_{\mathrm{f}}$ and $T_{\mathrm{g}}$ were found in the temperature dependence of the magnetization curve. 
The spontaneous magnetization appears below $T_{\mathrm{M}}$ and increases with decreasing temperature down to $T_{\mathrm{t}}$. 
Below $T_{\mathrm{t}}$ it decreases steeply down to approximately zero at around $T_{\mathrm{f}}$, at which it starts to increase again. 
The temperature dependence of the spontaneous magnetization shows a small drop at $T_{\mathrm{g}}$, above which the thermal conductivity shows a glassy behavior. 
The magnetic susceptibility shows the peak at $T_{\mathrm{N}}$, which is also seen in temperature dependence of magnetization under ZFC condition, although it was estimated by the magnetization curve measured under FC condition.

\ack{Acknowledgements}
This study was supported in part by a Grant-in-Aid for Scientific Research (A) (22246083) from MEXT and by the Japan Society for the Promotion of Science (JSPS) through its \textquotedblleft Funding Program for World-Leading Innovative R\& D on Science and Technology (FIRST Program).\textquotedblright

\end{document}